\documentclass[11pt,twoside]{article}
\usepackage{asp2010}

\resetcounters

\begin{document}

\title{Digital image exploration at Maui Community College}
\author{Katie M. Morzinski$^{1,2}$, Christopher J. Crockett$^{3,4}$, and Ian J. Crossfield$^{3}$}
\affil{$^1$Center~for~Adaptive~Optics, University~of~California~at~Santa~Cruz, 1156~High~St., Santa~Cruz, CA 95064}
\affil{$^2$Astronomy~Dept., University~of~California~at~Santa~Cruz, 1156~High~St., Santa~Cruz, CA 95064}
\affil{$^3$Astronomy~Dept., University~of~California~at~Los~Angeles, 430~Portola~Plaza, Box~951547, Los~Angeles, CA 90095}
\affil{$^4$Lowell~Observatory, 1400~W.~Mars Hill~Rd., Flagstaff, AZ 86001}

\begin{abstract}
  We designed a two-day laboratory exploration of fundamental concepts
  in digital images for an introductory engineering course at Maui
  Community College.  Our objective was for the students to understand
  spatial vs. brightness resolution, standard file formats, image
  tradeoffs, and the engineering design cycle.  We used open
  investigation, question generation, and an engineering design
  challenge to help our students achieve these learning goals.  We
  also experimented with incorporating Hawaiian
  language and cultural awareness into our activity.  We present our
  method, student response, and reflections on the success of our
  design.  The 2008 re-design of this activity focused on better
  incorporating authentic engineering process skills, and on using a rubric for
  summative assessment of the students' poster presentations.  A
  single file containing all documents and presentations used in this
  lesson is available
  online\footnote{\url{http://www.astro.ucla.edu/~ianc/files/digital_images_inquiry.pdf}}.
\end{abstract}

\section{Introduction}

In inquiry-style laboratory activities, students learn science
by performing science \citep{what_is_inquiry}.
Keys to inquiry are ownership of students over their learning
and authenticity of the activity to real-life science and engineering practices \citep{elements_of_inquiry}.
Here we discuss an engineering inquiry on Digital Image Files
we developed under the auspices of the
Professional Development Program (PDP).
The PDP is a unique educational program
that trains science, technology, engineering, and math (STEM) graduate students
to teach science and engineering
while simultaneously promoting STEM education at the undergraduate level
and for historically underrepresented populations \citep{PDPdesc}.
The PDP originated as part of the education theme of the National Science Foundation
Center for Adaptive Optics (CfAO),
and has now transformed to become a major component of the Institute for
Scientist and Engineer Educators (ISEE, Hunter~et~al., this volume).

\section{Activity Description}

\subsection{Venue Background}
ISEE is a key player in the Akamai Workforce Initiative (AWI),
a consortium that is developing education and employment opportunities for residents of the Hawaiian islands.
The Hawaiian word \textit{akamai} translates to \textit{clever},
and the goals of AWI are to
develop effective teaching in post-secondary schools in Hawai`i,
train local students for Maui-based careers in the technology industry,
increase the representation of women and Native Hawaiians in Hawai`i-based employment,
and build partnerships between high-tech educators and employers on Maui.

AWI encompasses internships, community programs, 
electro-optics certification, and curriculum development.
Curricula developed by the Teaching and Curriculum Collaborative (TeCC)
have provided support for creating a Bachelor's degree in Applied Science in Engineering Technology
at Maui Community College (MCC),
allowing the school to seek accreditation
as a four-year college---the University of Hawai`i, Maui College.
Toward this goal, in Fall 2008,
three TeCC teams were invited to design curricula for a new course,
\textit{Electronics 102: Instrumentation},
taught by MCC professor Mark Hoffman.
The TeCC teams designed three inquiries covering aspects of instrumentation:
CCDs (Mostafanezhad et al., this volume), Spectroscopy, and Digital Image Files.
This paper describes the Digital Images inquiry.

\subsection{Goals for Learners}
To plan this activity we first decided what we wanted the students to
get out of the experience.  We had four types of goals for the
students: content, process, attitudinal, and CfAO programmatic goals.
Our goals are summarized briefly in Table~\ref{tab:goals}.

\begin{table}[htbp]
  \caption{
    The learner goals we set out as we began the activity-planning process.
  }
  \smallskip
  \centering
  \begin{tabular}{ll}
    \hline
    {\bf Content Goals} & {\bf Process Goals} \\
    \hline
    Pictures can be represented by numbers     & Defining a problem              \\
    Pixels and arrays                          & Proposing a solution             \\
    Continuous vs. discrete                    & Communicating in writing\\
    Number of pixels and spatial resolution    &Evaluating tradeoffs                \\
    Bit depth and color resolution             & Solving a problem with constraints  \\
    Relation between file size and resolution & Carrying out engineering process \\
    Image file manipulation                    & \\
    Image file formats and header information  &\\
    \hline
    {\bf Attitudinal Goals} & {\bf CfAO Program Goals} \\
    \hline
    Solving a problem in a team  & Drawing on prior knowledge \\
    Being creative & Observing and communicating \\
    Making predictions            & Gaining career preparation \\
    Comfort in solving an engineering problem            &  \\
    \hline
  \end{tabular}
  \label{tab:goals}
\end{table}

\subsection{Overview of Activity}

We taught this activity at Maui Community College in Professor Mark
Hoffman's Electronic Instrumentation course to approximately 25 first-
and second-year students majoring in Electrical Engineering Technology.
The bulk of the hands-on investigation
encompassed image encoding and then decoding.  Students were provided
with astronomical images on paper, a light box, and a photometer.
Using these materials, students encoded their images into numbers.
Students then wrote up their encoded images into image files, swapped
them with other teams, and decoded a team's image by drawing the image
(with chalks) from looking at the image file.  Finally, students moved
to the computer lab to experience more in-depth digital image
manipulation.  Table~\ref{tab:timeline} shows the activity timeline;
we discuss the activity components in more detail below.

\begin{table}[htbp]
  \caption{
    The top-level schedule that we used in our  digital images inquiry.
  }
  \smallskip
  \centering
  \begin{tabular}{lclc}
    \hline
    \multicolumn{2}{l}{\textbf{Day 1}} & \multicolumn{2}{l}{\textbf{Day 2}} \\
    \hline
    Intro to Culture of Communication & 10 min. &     Intro: Day 2 & 10 min.\\
    Intro to Inquiry & 5 min.  &   Focused Investigation: & 30 min. \\
    Intro to Digital Images & 10 min. & \multicolumn{2}{c}{Image Decoding} \\
    Starters & 40 min.     & Prep for Sharing & 15 min.\\
    Break & 15 min. &    Sharing (``Jigsaw'') & 30 min.\\
    \multicolumn{2}{c}{(Facilitators sort questions)} &     Move to computer lab & 15 min.\\
    Starters Mini-Synthesis & 15 min. &   Image Manipulation & 30 min.\\
    Focused Investigation: &  60 min.  &  Discussion: & 10 min.\\
    \multicolumn{2}{c}{Image Encoding/Digitization} & \multicolumn{2}{c}{Communication experienced} \\
    Homework Assigned & 10 min.   &  Synthesis \& Closing & 25 min.\\
    \hline
     &  \textit{Total Time} & \textit{5 hrs.} & \\
    \hline
  \end{tabular}
  \label{tab:timeline}
\end{table}

\subsection{Activity Description}

\subsubsection{Starters}
A ``Starter'' is a brief, interactive pedagogical tool designed to
stimulate student interest and engagement in a topic, and to present
material relevant to subsequent components of an activity.
We used four Starters, rotating all the students through each one in parallel.
Each Starter was designed to introduce the students
to a particular concept relevant to our lesson goals.
We named our Starters
``Photometer Playground,''
``Flag Reproduction,''
``Pixels and Grayscale,''
and ``File Formats.''
An instructing facilitator was assigned to each Starter station.
After each Starter station, students wrote down their
questions, comments, and observations about that station;
these writings were collected by the activity facilitators for later discussion.

\begin{figure}[htbp]
\plottwo{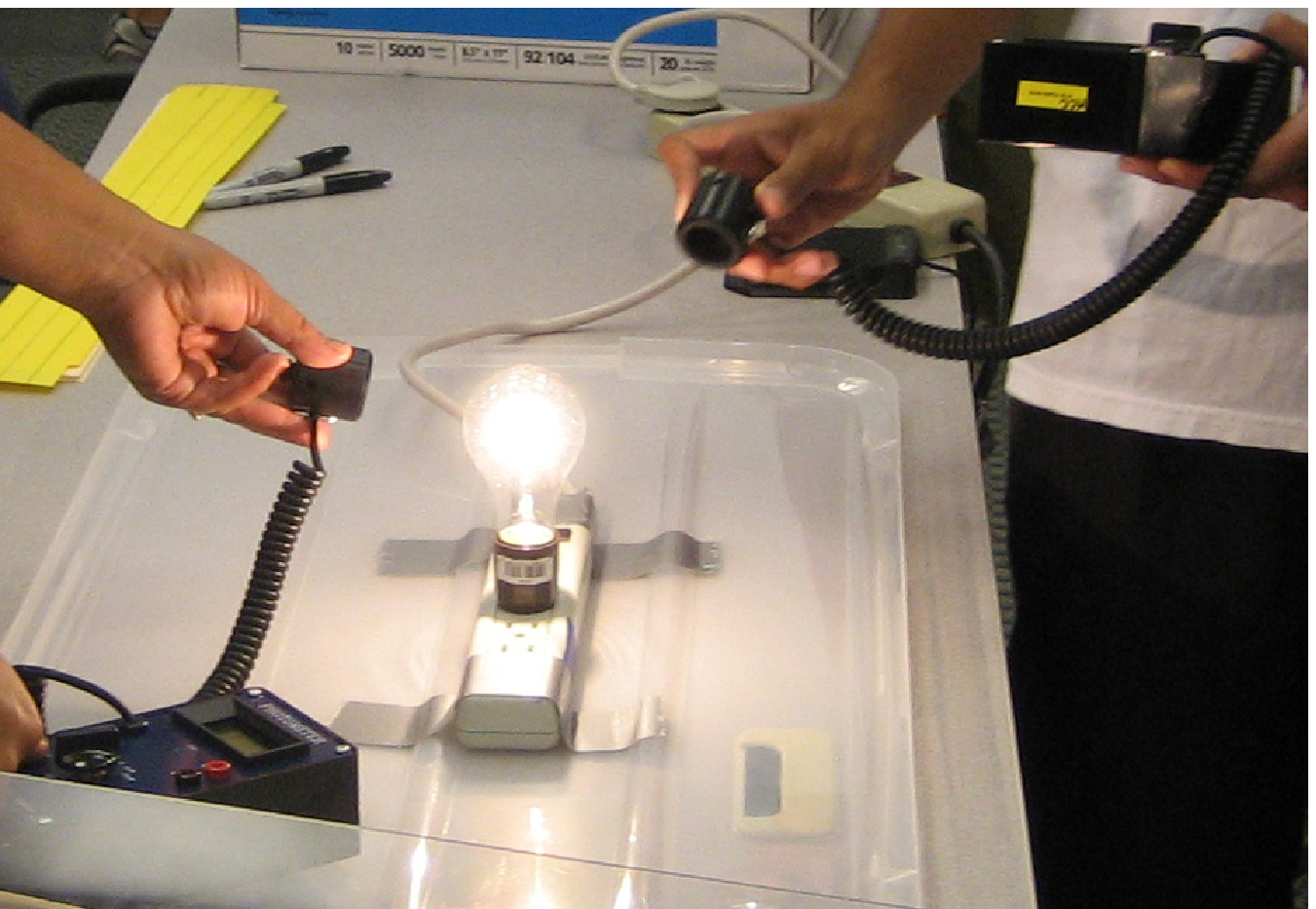}{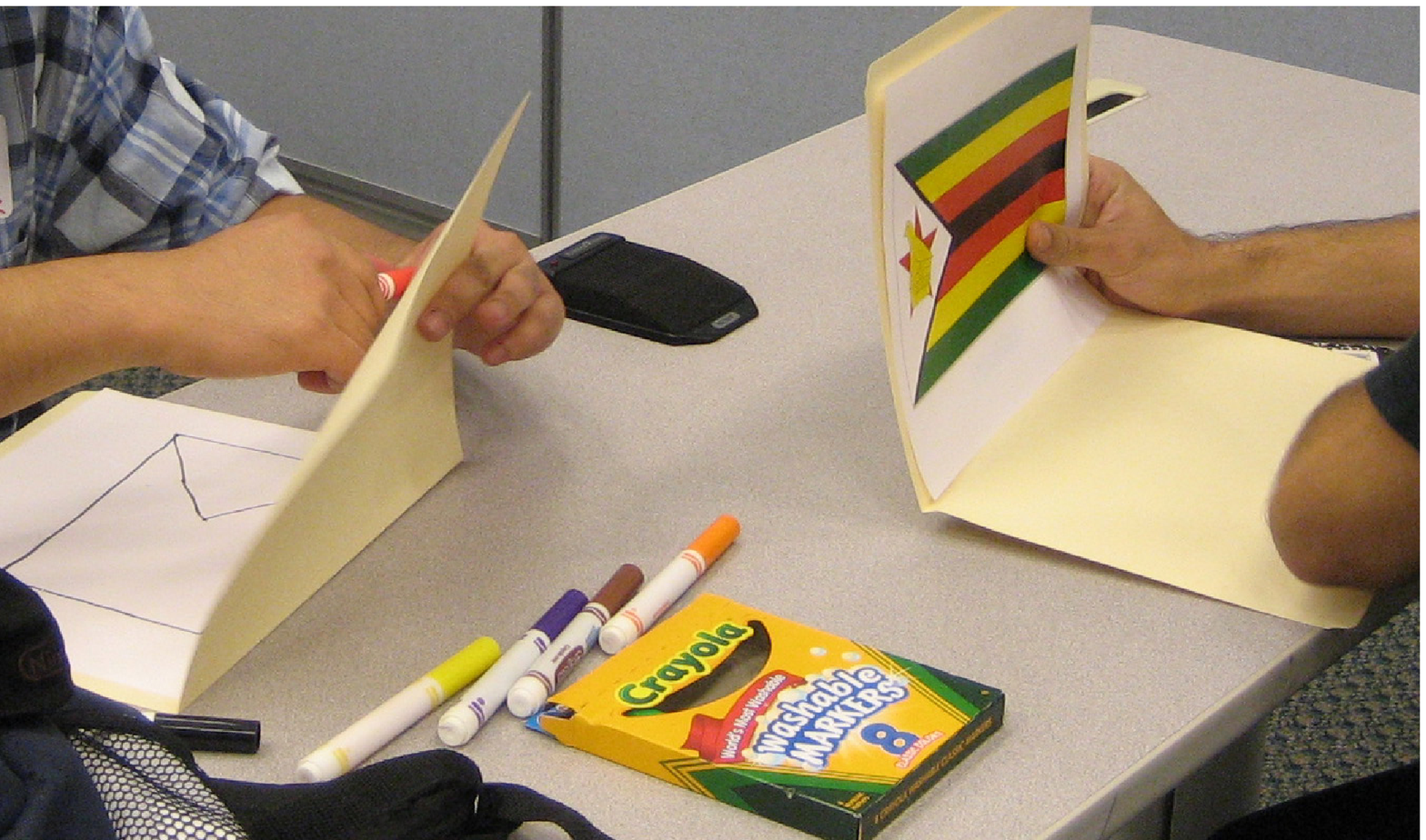}
        \caption{
                \label{fig:phot_flags}
                Starters: Using photometers to measure brightness in ``Photometer Playground'' (left)
                and transmitting an image verbally in ``Flag Reproduction'' (right).
                }
\end{figure}

At the ``Photometer Playground'' we introduced students to the use of
a photometer for measuring the intensity of incident light; this tool
was an essential component for the Focused Investigation that
followed. Students explored the use of a photometer to understand how
brightness can translate into a number.  They first observed 40-Watt,
100-Watt, and 300-Watt bulbs, and then explored the effects of
distance from and projection angle relative to the light source on
the photometer reading (see Figure \ref{fig:phot_flags}, left).
Finally, students observed the
photometer measurement when
attenuating the light through
paper printed with large two-inch squares of
white, gradations of gray, and black ink.
This was to demonstrate that a grayscale image
could be captured by shining a light through it and measuring the brightness with a photometer.

The purpose of ``Flag Reproduction'' was to encourage students to think
about how picture information can be communicated.
Students were paired off.
One member of each pair had a printed picture of an international flag
(chosen for a recognized format and simple geometric shapes).
The other member of the pair had a blank sheet of paper and colored markers.
Hiding the blank page and the flag from each other, the first student
described (verbally) the flag such that the second student could draw it
(see Figure \ref{fig:phot_flags}, right).
After doing their best to reproduce the flag,
students viewed the result and reflected on the process.

\begin{figure}[htbp]
\plotone{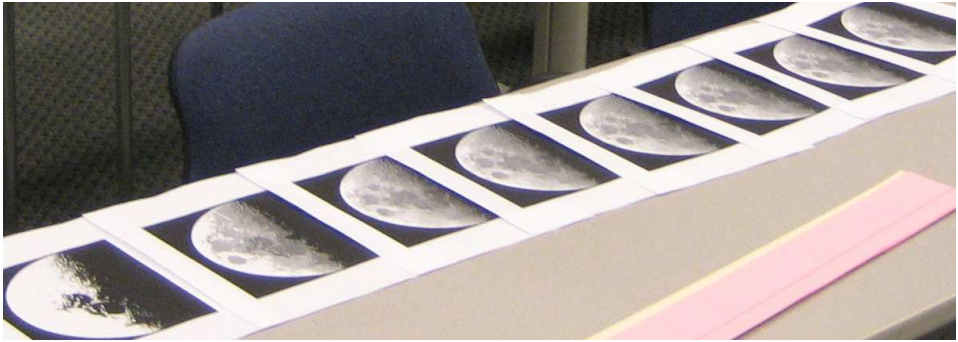}
        \caption{
                \label{fig:moon}
                Comparing images at different resolution in ``Pixels and Grayscale.''
                }
\end{figure}

The next Starter, ``Pixels and Grayscale,'' introduced students to the
ideas of pixel scale and bit depth (grayscale).  A grayscale
photograph of the moon was reproduced with ten varying pixel scales
and ten varying bit depths (see Figure \ref{fig:moon}).
The twenty images were arranged face-up
on a table, and students examined them and wrote questions or
observations.  Students were prompted to think of the differences
between the images, and advantages and disadvantages of each way of
representing the moon.

\begin{figure}[htbp]
\plotone{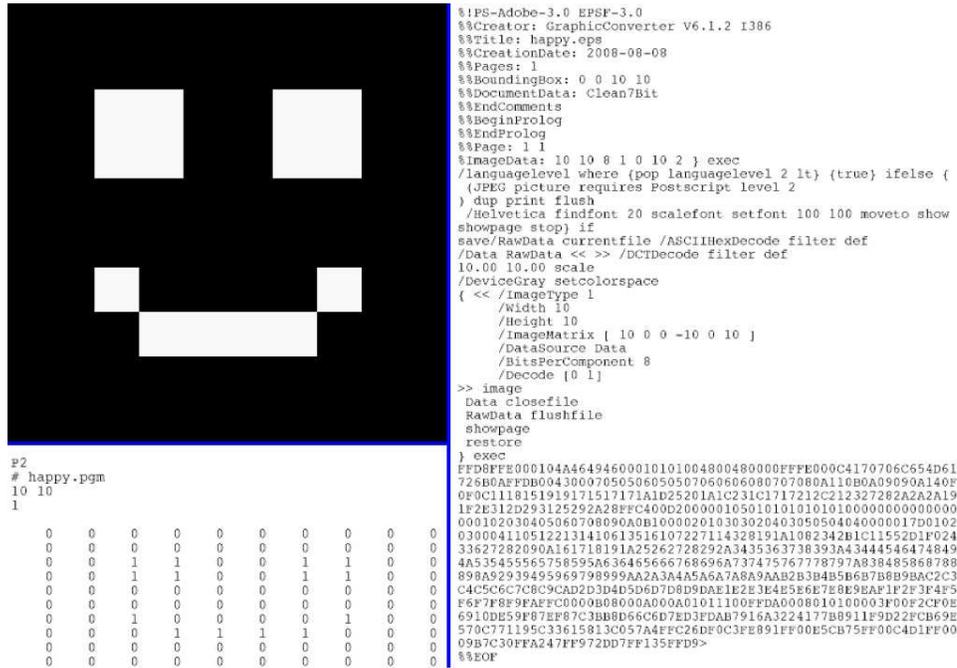}
        \caption{
                \label{fig:happy}
                Image (top left) and file data for the .pgm (lower left) and .eps (right) formats
                of the image.
                This was to illustrate representation of the same image in many different file formats
                 in the ``File Formats'' Starter.
                }
\end{figure}

We designed the fourth Starter, ``File Formats,'' to start students
thinking about how images are recorded in digital formats.  Students
were presented with one simple image (a black and white pixellated
``happy face'') encoded in a variety of formats (.eps, .fits, .jpg,
.pgm, .png, .svg).  The ASCII or hex data in each file was printed out
on the back of each image page,
and students were prompted to compare the pictures (which all looked the same)
and the ASCII/hex file formats, including both the header and body of the file formats (see Figure \ref{fig:happy}).
Prompts asked students to think about the differences and the
advantages and disadvantages of each.

\begin{table}[htbp]
  \caption{
    An edited sampling of the questions and observations generated by students during the Starters,
    sorted into categories corresponding to learning goals.
  }
  \smallskip
  \centering
  \begin{tabular}{l}
    \hline
    \textbf{Transmitting images}\\
    How do you communicate scale within a flag?\\
    Less data = easier to transmit/process\\
    \hline
    \textbf{Measuring light levels with a photometer}\\
    I notice the measurement gets smaller when the photometer is farther away\\
    The darker the sheet of paper through which the light goes, the lower the reading\\
    How much would turning off the room lights change the readings?\\
    \hline
    \textbf{Evaluating tradeoffs}\\
    Is there a sweet spot between good enough quality and too big of a file size?\\
    How many megapixels are needed for a sharp and clear image?\\
    Is there an advantage in using a short picture format vs. a long one?\\
    Some file formats are easier to be read by a human.  Are these not as useful?\\
    \hline
    \textbf{Information content}\\
    I believe that each pixel has its own number that represents its number in grayscale\\
    The image quality is not clear with limited pixels\\
    The more pixels there are, the overall quality of the images gets better and better\\
    \hline
    \textbf{Image file formats}\\
    Why are there so many different file types?\\
    Each one is formatted differently but all of them appear to be the same image\\
    The compressed formats JPEG, PNG are unreadable\\
    Which format will produce the best quality image?\\
    \hline
  \end{tabular}
  \label{tab:questions}
\end{table}

\begin{figure}[htbp]
\plotone{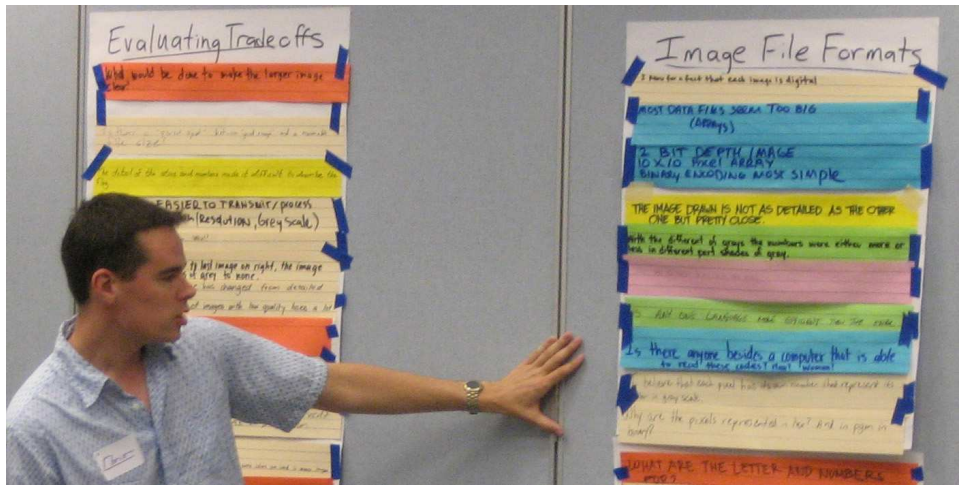}
        \caption{
                \label{fig:ques}
                Mini-synthesis of the students' observations from the Starters.
                }
\end{figure}

Table~\ref{tab:questions} lists a sample of the questions and observations
generated by students during the Starters. The questions generated by
students in our Starters were not used directly for the Focused
Investigations.  Rather, the questions were used to engage their
curiosity and introduce students to some of the concepts they would be
exploring later.  Students did not choose questions to investigate:
instead, the investigations were built around a particular engineering
challenge with images and digital image files.  Therefore, after the
break, we did a mini-``synthesis'' of the Starters (Figure \ref{fig:ques}) by going over the
questions generated with the students to ensure the knowledge gained
in the Starters became a shared classroom experience.

\subsubsection{Focused Investigation}

\begin{figure}[htbp]
\plotone{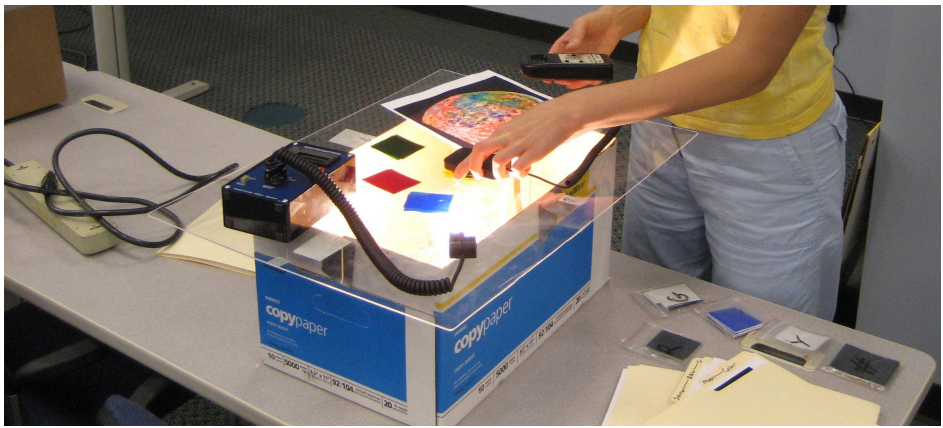}
        \caption{
                \label{fig:lightbox}
                Light box:
                An open box supports a sheet of plexiglass.
                Inside the box, a light bulb illuminates the image placed on top of the plexiglass.
                Measurements of the image brightness across the picture are made with a photometer.
                }
\end{figure}

For the Focused Investigation, students were given a grayscale
astronomy-related photograph.  Each team was also given an engineering
challenge in the form of various ``science cases'' to stimulate
different approaches.  The goals focused either on optimizing spatial
or color (grayscale) resolution, as explained in Table~\ref{tab:investigation}.
Furthermore, teams were given a
limited budget and a formula for the ``transmission cost'' per pixel
and color bit.  Their budget was \$1000 and pixels were \$2 each while
colors were \$50 each.  This ensured that teams could not maintain the
fidelity of the image in terms of both spatial and color resolution,
but rather had to make a tradeoff.
In anticipation of this, the ``Pixels and Grayscale'' Starter got students
to think about information content and number of pixels or color bits
in an image.
Of course, students were also
limited by the limited amount of time they had to use the photometers
-- not all groups considered this during their planning!  Students
were told to record the image using letters and numbers only, so that
they could transmit the image to another team who would then re-create
the image with the goal as given, for example mapping sunspots.  They
used the photometers to do so, digitizing their images by hand
using the light boxes (Figure \ref{fig:lightbox}) as practiced in the ``Photometer Playground'' Starter.

 \begin{table}[htbp]
   \caption{
   Goals for encoding each image during the Focused Investigation.
   Each team had one image and one goal, focused on either spatial or color resolution.
   }
   \smallskip
   \centering
   \begin{tabular}{lll}
     \hline
     \textbf{Image} & \textbf{Spatial Resolution Goal} & \textbf{Color Resolution Goal}\\
     \hline
     Sun & Differential rotation rate & Temperature of sunspots\\
      & \textit{Map sunspots in time} & \textit{Brightness of sunspots}\\
     \hline
     Moon & Elevation topography & Temperature of rocks\\
                & \textit{Map maria \& terrae} & \textit{Brightness of rocks}\\
     \hline
     Jupiter & Rotation period & Height of clouds\\
                & \textit{Map clouds in time} & \textit{Brightness of clouds}\\
     \hline
     Saturn & Ring structure & Chemical composition\\
                & \textit{Map rings} & \textit{Brightness of atmosphere}\\
     \hline
   \end{tabular}
   \label{tab:investigation}
\end{table}

For homework after the first 2.5-hour course session, students had to
write up their digitized image along with a file format description
(to provide directions on how to decode their image data).  On Day 2,
students swapped images and used the written information to re-draw
the image using grayscale chalks.  This was an exercise anticipated in the
``Flag Reproduction'' Starter when students practiced communicating an image
and in the ``File Formats'' Starter when students were exposed to 
different file formats for describing the same image.

\subsubsection{Sharing}

\begin{figure}[htbp]
\plottwo{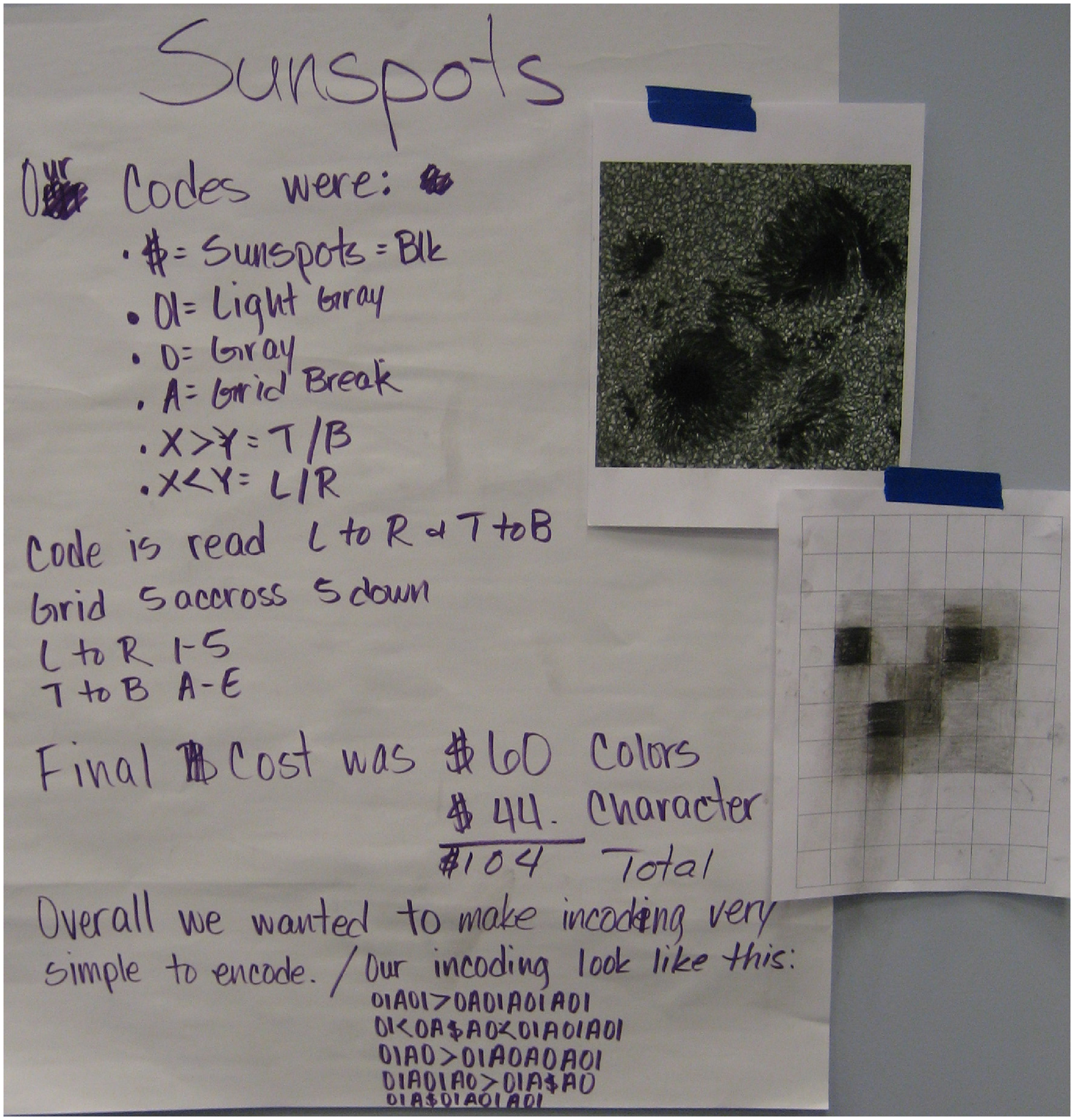}{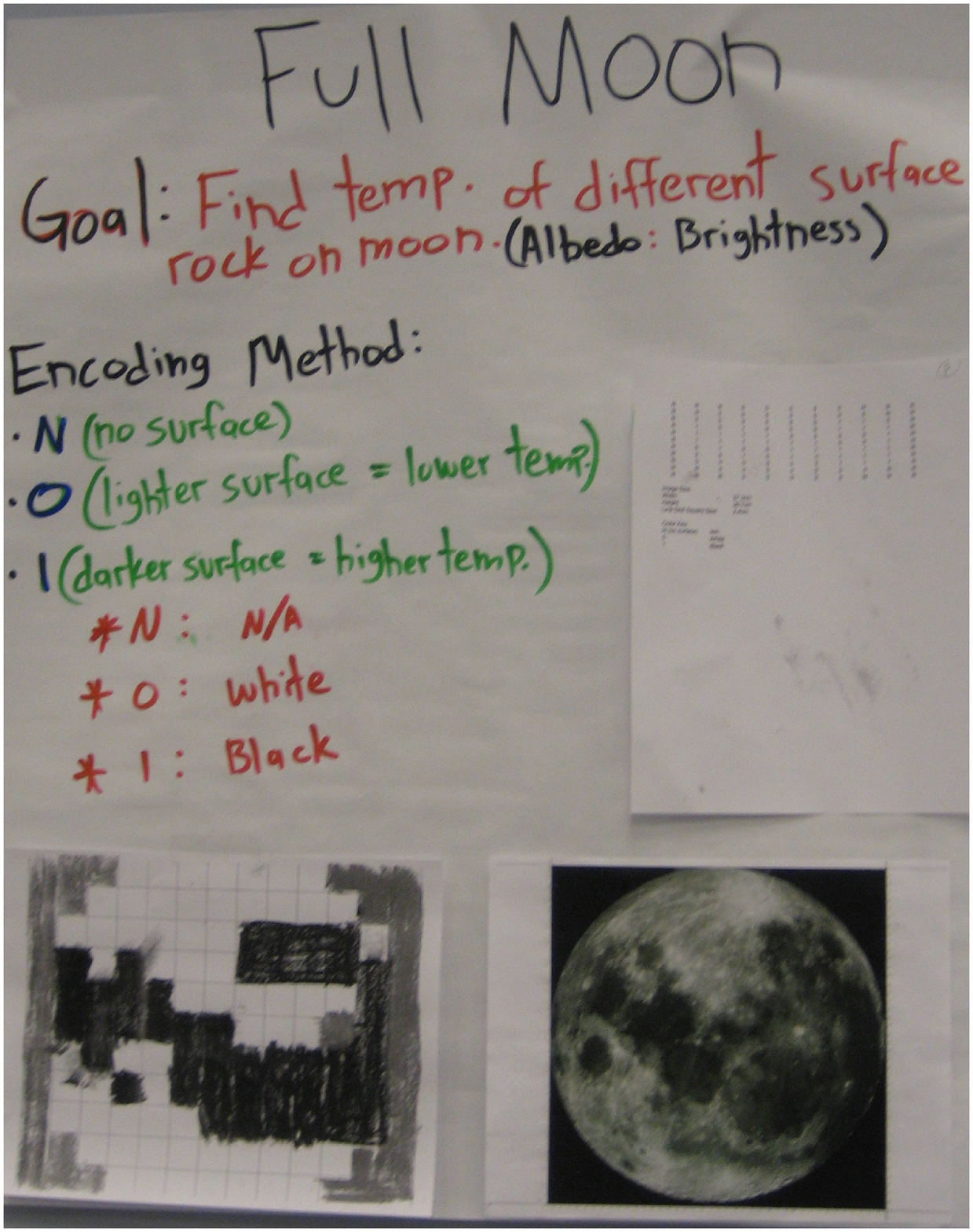}
        \caption{
                \label{fig:posters}
                Posters by students for sharing.
                }
\end{figure}

After each team had reproduced another team's image file,
the decoded drawings were handed back to the original team for sharing.
Facilitators photocopied the drawings so that each student would have a copy.
We used a ``Jigsaw''-style sharing in which each team split up and sent
one team member to each facilitator to share their results to one-third of the class.
This ensured that each student was responsible for all the material.
Students made posters stating their science goal from Table~\ref{tab:investigation},
 describing their tradeoffs in encoding or digitizing their image,
 displaying the resulting drawing,
 and reflecting on the investigation.
Figure~\ref{fig:posters} shows two students' posters.
  
Students presented individually to one of the three facilitators, and facilitators scored their presentations with a rubric (Table~\ref{tab:rubric}) as a tool to conduct a summative assessment of the students' learning.
Our design team was one of the first PDP design teams to pilot use of a rubric for inquiry.
We chose three categories on which to grade each presentation: describing the encoding process,
describing the image file, and practicing good communication skills.
We expected students' level of mastery to advance in proficiency from left to right along a row in the rubric.  However, to allow for a student achieving mastery at the last column along a given row yet missing one of the more basic items in another row, we awarded students 1 point per cell.

\begin{table}[htbp]
  \caption{
  Rubric used for summative assessment.
  }
\small
  \smallskip
  \centering
  \begin{tabular}{m{0.01cm}m{1.6cm}m{3.0cm}m{3.0cm}m{3.0cm}m{0.01cm}}
    \hline
    & \textbf{Task} & \raggedright{\textbf{Did not meet expectations [+1]}} & \raggedright{\textbf{Met expectations [+1]}} & \raggedright{\textbf{Exceeded expectations [+1]}} & \\
    \hline
    & \raggedright{\textbf{Describe team's image encoding process and choices.}} & \raggedright{Student shows their original image, the drawing another team made of it, and explains the scientific goal they were working toward.} & \raggedright{Student describes their image encoding method (photometer digitization, vector graphics, or other).} & \raggedright{Student explains the tradeoffs they evaluated and gives reasons for choosing their image encoding method.} & \\
    \hline
    & \raggedright{\textbf{Describe team's image file format, giving reasons.}} & \raggedright{Student shows their image file format, identifying the header and body.} & \raggedright{Student explains what the header and body mean, and why the particular image file format was chosen to meet the scientific goals.}  & \raggedright{Student evaluates the clarity of their image file format by the fidelity of the drawn image, and suggests changes they could have made to clarify their image encoding.}  & \\
    \hline
    & \raggedright{\textbf{Show communication skills.}} & \raggedright{Student speaks and has visual aids.} & \raggedright{Student speaks clearly and audibly, and has visual aids that are legible and appropriate.}  & \raggedright{Student engages in relevant discussion with classmates about presentation.}  & \\
    \hline
  \end{tabular}
  \label{tab:rubric}
\end{table}
\normalsize

\subsubsection{Computer lab}

After sharing what students had learned in digitizing and transmitting images by hand,
we moved to the computer lab to do an exercise with images in the .pgm format.
Students manipulated the numbers in a simple .pgm image of the moon, and then viewed
the results with the image displaying program \textit{Irfanview}.
We provided students with prompts such as making the image darker
or inverting the colors.
This exercise reinforced the idea that digital images are represented by numbers in arrays
and that the values in the image body represent the brightness of each pixel.

\subsubsection{Closing}

Finally, we wrapped up the lab with a reflection on the different ways
communication expert Kalei Tsuha of MCC had
observed students communicating throughout the activity,
followed by a synthesis lecture of what students had learned.
For homework, students were asked to produce a report justifying their decisions
in light of their constraints and science goals.
In this report, the students were expected to discuss the possible design tradeoffs,
the limitations of their design, and how they might redesign their solution in the future.
 
\section{Discussion}

We asked the students to fill out written feedback forms to improve our instruction in the future,
and some concepts students wanted to explore further included more practice encoding or digitizing images, more on image formatting and compression, and more on image manipulation.
Students rated each component of the activity on a five-point scale and results are shown
in Table \ref{tab:feedback}.
Students got the most out of the image decoding, poster sharing, and synthesis lecture.

\begin{table}[htbp]
  \caption{
  Student feedback on a five-point scale.
  }
  \smallskip
  \centering
  \begin{tabular}{lcc}
    \hline
    \textbf{Activity Component} & \textbf{Mean Score} & \textbf{Std. Dev.} \\
    \hline
    Starter & 3.9 & 1.2 \\
    Image Encoding & 3.9 & 1.3 \\
    Homework: File Creation & 3.6 & 1.4 \\
    Image Decoding & 4.4 & 0.8 \\
    Poster Session & 4.5 & 0.8 \\
    Computer Activities & 3.9 & 1.3 \\
    Synthesis Lecture & 4.5 & 0.6 \\
    \hline
    \end{tabular}
  \label{tab:feedback}
\end{table}

This activity in Fall 2008 was a redesign of a similar Digital Image Files inquiry taught in Spring 2008.
In the redesign we attempted to add more authenticity to the engineering challenge
by both tying it to a science goal
(e.g., a goal of mapping sunspots to motivate a focus on optimizing spatial resolution)
as well as the monetary budget constraint.
We got feedback from the course instructor Elisabeth Reader, in reviewing the write-up
assigned on Day 2, that many students still found identifying tradeoffs to be difficult.
Upon reflection, the budgetary constraint may have been too complicated,
and it the future we would like to spend more time on clarifying the science goals in Table~\ref{tab:investigation}
so that students can better make tradeoffs to optimize achievement of the goal
in a more authentic way.

On the whole, the inquiry was a success as students learned about digital images,
pixels, transmitting and communicating images, and making tradeoffs.

As inquiry designers and facilitators, we feel we accomplished our goals in this activity.
After the effort involved in designing and teaching,
we would be pleased to see our work go farther and have thus made all the materials
and a lesson plan available on the website of facilitator
IJC\footnote{\url{http://www.astro.ucla.edu/~ianc/files/digital_images_inquiry.pdf}}.
MCC instructor Elisabeth Reader has already taught the activity again
with a new class, also finding it successful.

\acknowledgements

Lisa Hunter was an observer and design-team consultant at MCC,
while Lynne Rashke was a consultant at the PDP workshop. 
UH-Maui professors Mark Hoffman and Elisabeth Reader (the classroom teachers hosting this inquiry) and John Pye provided advice and assistance.
Hawaiian language and culture expert Kalei Tsuha consulted
and contributed to the culture and communication portion.
J.~D.~Armstrong and Joe Masiero were PDP participants who designed the in\-i\-tial Digital Images inquiry
at the 2007 PDP and taught the inquiry as a pilot in Spring 2008.

This material is based upon work supported by: the National Science Foundation (NSF) Science and Technology Center program through the Center for Adaptive Optics, managed by the University of California at Santa Cruz (UCSC) under cooperative agreement AST\#9876783; NSF AST\#0836053; NSF AST\#0850532; NSF AST\#0710699; Air Force Office of Scientific Research (via NSF AST\#0710699); the UCSC Institute for Scientist \& Engineer Educators; and the University of Hawai`i.

\bibliography{morzinski}

\end{document}